\documentclass{aastex}
\usepackage{emulateapj5}



\def\chandra    {{\em Chandra}\/}
\def\xmm        {XMM-{\em Newton}\/}
\def\rosat      {{\em ROSAT}\/}

\begin{document}

\title{The pressure profiles of hot gas in local galaxy groups}

\author{
M.\ Sun\altaffilmark{1},
N.\ Sehgal\altaffilmark{2},
G.\ M.\ Voit\altaffilmark{3},
M.\ Donahue\altaffilmark{3},
C.\ Jones\altaffilmark{4},
W.\ Forman\altaffilmark{4},
A.\ Vikhlinin\altaffilmark{4},
C.\ Sarazin\altaffilmark{1}
}

\altaffiltext{1}{Department of Astronomy, University of Virginia, P.O. Box 400325, Charlottesville, VA 22904-4325; msun@virginia.edu}
\altaffiltext{2}{Kavli Institute for Particle Astrophysics and Cosmology, Stanford University, Stanford, CA, USA 94305-4085}
\altaffiltext{3}{Department of Physics and Astronomy, MSU, East Lansing, MI 48824}
\altaffiltext{4}{Harvard-Smithsonian Center for Astrophysics,
60 Garden St., Cambridge, MA 02138}

\shorttitle{Group pressure profiles}
\shortauthors{SUN ET AL.}

\begin{abstract}

Recent measurements of the Sunyaev-Zel'dovich (SZ) angular power spectrum from
the South Pole Telescope (SPT) and the Atacama Cosmology Telescope (ACT)
demonstrate the importance of understanding baryon physics
when using the SZ power spectrum to constrain cosmology. This is challenging
since roughly half of the SZ power at $\ell$=3000 is from low-mass systems with
$10^{13} h^{-1}$ M$_{\odot} < M_{500} < 1.5\times10^{14} h^{-1}$ M$_{\odot}$,
which are more difficult to study than systems of higher mass.
We present a study of the thermal pressure content for a sample of local galaxy
groups from Sun et al.~(2009).  The group $Y_{\rm sph, 500} - M_{500}$ relation
agrees with the one for clusters derived by Arnaud et al.~(2010).
The group median pressure profile also agrees with the universal pressure profile
for clusters derived by Arnaud et al.~(2010).
With this in mind, we briefly discuss several
ways to alleviate the tension between the measured low SZ power and the
predictions from SZ templates.

\end{abstract}


\section{Introduction}

Galaxy groups are not scaled-down versions of rich clusters following simple
self-similar relations (e.g., Ponman et al. 2003; Voit 2005).
They are systems where the role of complex baryon physics (e.g., cooling,
galactic winds, and AGN feedback) begins to dominate over gravity. The effects of these
baryon processes are not large but still significant in massive clusters, and therefore
need to be calibrated if we want to further improve the cosmological constraints from
clusters. Since the role of these processes is less pronounced in massive clusters,
it is easier to study and understand them by observing groups.
The importance of galaxy groups for cosmology has also been well demonstrated by the recent
measurements of the Sunyaev-Zel'dovich (SZ) angular power spectrum. 
SPT measured a value for the SZ power at $\ell=3000$ (scales of $\sim 4'$) that was lower
than most prior predictions by at least a factor of two (Lueker et al. 2010).
A measurement of the SZ power spectrum by ACT is consistent with these results
(Das et al.~2010; Dunkley et al. 2010).
The SZ angular power spectrum is a sensitive probe both of cosmological
parameters and of the hot gas content of galaxy clusters and groups (e.g. Komatsu \& Seljak 2002).
Regarding the latter, it opens a new window into low-mass, high-redshift systems as about
half of the SZ power at $\ell$=3000 comes from halos with
$10^{13} h^{-1}$ M$_{\odot} < M_{500} < 1.5\times10^{14} h^{-1}$
M$_{\odot}$ and $z > 0.5$ (e.g., Trac et al.~2010).
While the examination of the thermal pressure content in $z > 0.5$ groups is a challenge to
current X-ray telescopes with typical exposures, such work can be done for local groups.
Group pressure profiles have received little attention to date and the existing samples
are small (e.g., Mahdavi et al. 2005; Finoguenov et al. 2007).
In this letter, we present the pressure profiles
of hot gas in 43 local galaxy groups from the Sun et al. (2009, S09 hereafter) sample.
We assume $\Omega$$_{\rm M}$=0.24, $\Omega_{\rm \Lambda}$=0.76 and
H$_{0}$ = 73 km s$^{-1}$ Mpc$^{-1}$.

\section{The group sample and data analysis}

The group sample and the \chandra\ data analysis have been discussed in S09. There are
43 groups at $z$=0.01 - 0.12 (a median $z$ of 0.033), all with intracluster medium (ICM) properties
derived to at least $r_{2500}$. Twenty-three have masses measured to $\sim r_{500}$.
The mass range is $M_{500} = 10^{13} h^{-1}$ M$_{\odot} - 10^{14} h^{-1}$ M$_{\odot}$.
As an archival sample, there is not a well defined selection function. The sample
does include some X-ray faint groups, usually with strong radio AGNs at the center.

The analysis in S09 was based on \chandra\ CALDB 3.4.3. Since then, there
have been major \chandra\ calibration releases on the on-axis effective area
and the ACIS low-energy contamination models.
To check the impact of calibration changes on the S09 results, we examined
spectra from 40 regions from 18 observations which we studied in S09. The temperatures of these
regions range from 0.7 to 3.0 keV and the abundances range from 0.15 to 1.5 solar.
The observation dates are from 2000 to 2006, and both ACIS-I and ACIS-S
observations are examined. CIAO 4.3 with CALDB 4.4.1 and XSPEC 12.6 were used.
The temperature decrease with the new calibration is less than 1\% on average.
This is not surprising (e.g., Nevalainen et al. 2010)
as temperatures at this range are mainly determined by the centroid of the iron-L hump.
The decrease of the ACIS effective area below 5 keV causes the normalization of
the spectral model to increase, on average, by 9.9\%, which is independent of the
temperature. This implies an average 4.9\% increase in gas density. We do not include
this small change in this work.

While small ACIS calibration uncertainties may still remain after the release
of CALDB 4.4.1, the results for groups should not be
affected much. The \chandra\ flux in the 0.5 - 2.0 keV band agrees with
the \xmm\ flux to better than 4\% (e.g., Nevalainen et al. 2010). The existing
temperature bias is small for low-temperature gas as long as the iron-L hump
is significant.
Even if temperatures of low-temperature gas are still biased high by
e.g., 10\%, the $Y_{\rm sph, 500} - M_{500}$ relation (Fig. 1) should not be
affected much as $M_{500}$ and $Y_{\rm sph, 500}$ will both be lower, by $\sim$ 15\%
and $\sim$ 20\% respectively, producing a small change in the relation that is insignificant
compared to the large statistical uncertainties on $M_{500}$
(a median 1$\sigma$ uncertainty of 18\%).

\section{The integrated pressure content}

We first compare the scaling relation between mass and the volume-integrated
Compton parameter for the S09 groups with the Arnaud et al. (2010, A10 hereafter)
results. The A10 sample is
from \xmm\ observations of 31 Representative \xmm\ Cluster Structure Survey (REXCESS) clusters
with $M_{500} = 7\times10^{13} h^{-1}$ M$_{\odot}$ - $6\times10^{14} h^{-1}$ M$_{\odot}$.
The spherically integrated quantity, $Y_{\rm sph, 500}$
(defined in Equ.~14 in A10), is derived within $r_{500}$.
The S09 $Y_{\rm sph, 500} - M_{500}$ relation agrees well with A10's (Fig. 1). 
This can be expected from the good agreement of the $M_{500} - Y_{\rm X, 500}$ relation
between the two works.
The difference between $Y_{\rm X, 500}$ and $Y_{\rm sph, 500}$ is the ratio
between the spectroscopic temperature and the gas-mass-weighted temperature within
$r_{500}$ ($T_{\rm X}$ vs. $T_{\rm mg, 500}$). The $T_{\rm X}$ in S09 is measured between
0.15 $r_{500}$ and $r_{500}$ (we call it $T_{500}$), while the $T_{\rm X}$ in A10 is measured between
0.15 $r_{500}$ and 0.75 $r_{500}$. For the S09 sample, $T_{\rm mg, 500}$ / $T_{500}$ =
0.98$\pm$0.04 and $T_{500}$ = (0.95$\pm$0.02) $T_{\rm X}$(0.15 $r_{500}$ - 0.75 $r_{500}$).
Arnaud et al. (2007) quoted 0.94-0.97 for the latter ratio.
Thus, $T_{\rm mg, 500}$ / $T_{\rm X}$(0.15 $r_{500}$ - 0.75 $r_{500}$) $\sim$ 0.93
for the S09 sample, which agrees with the A10 result of 0.924$\pm$0.004. 
Because of this consistency and the agreement between the $M_{500} - Y_{\rm X, 500}$
relations, we expect the $Y_{\rm sph, 500} - M_{500}$ relations from S09 and A10 to agree.
We also examined the twenty tier 3 and 4 groups in S09 (where gas properties are only derived to
45\% - 72\% of $r_{500}$). The $Y_{\rm sph, 2500}$ - $T_{500}$
relation for the tier 3 and 4 groups agrees with the relation for the tier 1 and 2 groups.
Overall, we conclude that for the S09 groups, the derived $Y_{\rm sph, 500}$
is 1.05$\pm$0.25 times $Y_{\rm sph, 500}$ predicted from the A10 relation.
We also plot the predicted $Y_{\rm sph, 500} - M_{500}$ relations from recent
SZ templates (Sehgal et al. 2010a; Battaglia et al. 2010; Shaw et al. 2010; Trac et al. 2010)
in Fig. 1. The Sehgal et al. (2010a) template used the ICM model by Bode et al. (2009),
which was calibrated with the Vikhlinin et al. (2006) and S09 gas fraction relations.
Vikhlinin et al. (2009) included six more clusters with on average lower gas fractions within
$r_{500}$ than the eight clusters in Vikhlinin et al. (2006). As the $M_{500} - Y_{\rm X,500}$
relation from Vikhlinin et al. (2009) agrees well with that in A10, it is not surprising
that the Sehgal et al. (2010a) template has $\sim$ 9\% higher normalization than the
A10 result for clusters.

\section{The radial pressure profiles}

We also examine the radial pressure profiles of the S09 groups.
A10 derived a universal pressure profile of the ICM by removing the mass dependence.
If the ICM scaling relations are self-similar, the mass dependence is $M^{2/3}$
(Equ.~5 of A10, $P_{500}$, also see Nagai et al. 2007). Since deviations
from self-similarity exist, A10 defined a term (Equ.~7 and 8 of A10, which we call
$P_{\rm adjust}$) to further remove the mass dependence in addition to $P_{500}$, where
$P_{\rm adjust} \propto M_{500}^{\alpha (x)}$ and
$\alpha (x$) = 0.22/(1+$x^{3}$), $x = 2 r/r_{500}$.
While the form of $P_{\rm adjust}$ can be examined with the S09
groups, the large uncertainties in $M_{500}$ for the S09 groups
do not allow good constraints on this adjustment factor. 
We derived the $P / P_{500}$ and the $P / P_{500} / P_{\rm adjust}$
profiles for all 43 groups. Uncertainties are estimated from those in the
temperature and density profiles. In Fig. 2, we show the median and the
1$\sigma$ scatter of both profiles. To account for uncertainties in
pressure profiles, 1000 Monte Carlo simulations were run to examine
the uncertainties on the median profiles, which is small as shown in Fig. 2.
The 1$\sigma$ scatter for $P / P_{500} / P_{\rm adjust}$ is 26\% - 40\% at 0.2 $r_{500}$ - 0.8 $r_{500}$
(compared to less than 30\% scatter for the A10 clusters) and is consistent with a log-normal form.
Thus, the radial pressure profiles of the S09 groups agree well with
the universal pressure profile defined by A10.
This is consistent with the good agreement between the two works on the mass
proxies and entropy scalings (S09; Pratt et al. 2010).

\section{Discussions and conclusions}

The results above suggest that the thermal pressure of local galaxy groups from the
S09 sample is consistent with the extrapolation from the A10 results, although
statistical errors and scatter are still large.
Interestingly, recent measurements of the SZ angular power spectrum are at
least a factor of two lower than prior expectations at $\ell=3000$ (Lueker et al.~2010).
The thermal SZ power spectrum scales roughly as the square of the thermal SZ flux.
Given the results presented above,
we briefly discuss several possibilities that may alleviate this tension.

{\bf X-ray selection bias:}
The S09 sample is an X-ray archival sample and the REXCESS sample is an
X-ray-luminosity-selected sample. Both samples can be different from
mass-selected samples. The \chandra\ archive may be biased to systems
with bright cores, while X-ray under-luminous groups and clusters may exist
(e.g., Rasmussen et al. 2006; Popesso et al. 2007).
However, the $Y_{\rm sph} - M$ relation at $r_{500}$
and beyond is less affected by the presence of X-ray bright regions
(e.g., a large cool core) than the $L_{\rm X} - M$ relation.
For the S09 sample, 12\% - 68\% of the X-ray flux (a median of $\sim$ 34\%)
is from within $0.15 r_{500}$, while such regions
only contribute $\sim$ 5\% to $Y_{\rm sph, 500}$ for
$M_{500} = 10^{13} h^{-1}$ - $10^{15} h^{-1}$ M$_{\odot}$ halos,
assuming the A10 pressure profile.
The contribution may be even smaller than 5\% for X-ray under-luminous
systems as their gas cores are fainter than those of the REXCESS clusters
used to derive the A10 profile.
One main conclusion of the S09 work is that the gas content of groups is comparable
to that of clusters at $r > r_{2500}$, at least for the S09 sample. 
If we combine the $n_{\rm e} - T_{500}$ relations from Vikhlinin et al. (2009),
S09 and REXCESS, the trend of slope flattening with increasing radius is
significant, with an almost constant density at $r_{500}$ from
groups to clusters.
This trend is consistent with the scenario that much of the
low-entropy gas in low-mass systems has been ejected to large radii by
strong feedback (e.g., McCarthy et al. 2010).
However, it remains to see whether this result applies to mass-selected samples.
One way to test this is to examine scaling relations from
non-X-ray-selected samples. This kind of work has been done on
optically selected samples by stacking the \rosat\ all-sky survey data
(Dai et al. 2010; Rykoff et al. 2008). Besides the systematic uncertainties with
stacking, the \rosat\ temperatures from stacking are often biased
(Rykoff et al. 2008) and contamination to those samples, especially
at the low-mass end, can be severe.

{\bf Pressure contribution at ${\bf r > r_{500}}$:}
The total SZ flux is more sensitive to the gas in cluster outskirts ($r > r_{500}$) than
the total X-ray flux. Few direct X-ray constraints exist at such large radii,
especially for groups.
Although the contribution from $r > r_{500}$ to $Y_{\rm sph}$ is significant
(e.g., 40\% - 70\% increase by integrating to 2$r_{500}$ for the A10 profile),
the contribution to the SZ power spectrum at $\ell$=3000 assuming the A10 profile
is smaller, only about 20\% from $r > r_{500}$ regions. So an overestimate of the
thermal pressure from the A10 model only at $r > r_{500}$ could overpredict the SZ
power spectrum by at most 20\% at $\ell$=3000.

{\bf Dynamical state of the ICM:}
The SZ signal measures the total thermal energy of electrons. However, the
potential energy of halos may not be fully converted into thermal energy
of electrons because
of e.g., recent mergers and weak viscosity of the ICM (e.g., Lau et al. 2009;
Burns et al. 2010). Galaxies can also contribute to the non-thermal
pressure support by e.g., injected magnetic fields and cosmic rays.
The non-thermal pressure support may also cause the ICM to be clumpy.
For a clumpy ICM, the SZ signal predicted from the X-ray data will be biased
high. All these effects may have a dependence on mass (or the ICM temperature),
and the evolution of these effects with redshift may not be self-similar.

The impact of the non-thermal pressure support on the SZ power spectrum has been
discussed by Shaw et al. (2010), who examined models with a radial dependence of
the non-thermal pressure.
Trac et al. (2010) examined a model with $20\%$ non-thermal pressure for
all clusters and groups at all masses and redshifts. 
However, both models do not predict the mean value measured for the SZ power spectrum at $\ell=3000$
by Lueker et al. (2010), being high by about 1$\sigma$.
Interestingly, SZ observations suggest the latter model predicts too little SZ
flux for very massive clusters (Sehgal et al.~2010b).
As for clumpiness, the good agreement between the measured SZ radial profile and the prediction
from X-ray data for individual clusters (e.g., Plagge et al 2010; Komatsu et al. 2010; Sehgal et al. 2010b)
suggests that clumpiness should be weak for massive clusters.
However, one can imagine a mass dependence for clumpiness, as both
heat conduction and dynamic viscosity can be much weaker in groups than in clusters.

If the hydrostatic equilibrium mass under-estimates the true mass by 20\%
from $M_{500}$ = 10$^{13}$ h$^{-1}$ M$_{\odot}$ - 10$^{15}$ h$^{-1}$ M$_{\odot}$
(e.g., Nagai et al. 2007),
$Y_{\rm sph, 500}$ will increase by 4\% - 7\%, if the universal pressure
profile from A10 is assumed. Therefore, the normalization of the
$Y_{\rm sph, 500} - M_{500}$ relation will decrease by $\sim$ 24\%,
which roughly translates to $\sim$ 42\% decrease on the predicted
SZ power spectrum.
If the mass bias is larger for low-mass halos, the relation will be steeper
and the decrease will be larger.
However, it is a big challenge to constrain the effects of non-thermal pressure
support in groups and clusters, especially its dependence with mass and
redshift. For groups, robust mass measurements that do not assume
hydrostatic equilibrium are required. Two promising methods
are stacking of the lensing data (e.g., Leauthaud et al. 2010) and caustics
(Rines \& Diaferio 2010).
Future X-ray microcalorimeter observations (e.g., by {\em Astro-H})
may also constrain the ICM turbulence directly. 

{\bf Evolution of the ICM properties:}
While local groups are discussed in this paper, most of the SZ power at $\ell$=3000
from groups is from systems at $z>0.5$. Evolution of the ICM properties is poorly
constrained for poor clusters and groups.
Recent results on $z \geq 0.5$ groups suggest that the evolution of the
$L_{\rm X}$ scalings is not weaker than the self-similar prediction
(e.g., Jeltema et al. 2009; Leauthaud et al. 2010), but the statistical uncertainties
are large.
Proper understanding of the evolution of the $L_{\rm X}$ scalings
requires a good understanding of the selection function (e.g. Pacaud et al. 2007). 
Better constraints on the evolution of the low-mass end of the $L_{\rm X}$ scaling relations
should be achieved with more \xmm\ and \chandra\ data on larger samples
with well-defined selection functions.
Of course, the evolution of the $L_{\rm X}$ scaling relations is not equal to
the evolution of the $Y - M$ relation. More factors, e.g., the evolution
of the cool core fraction and the gas distribution, need to be accounted
for. Alternatively, deep SZ observations of high-$z$ groups, either
individually or by stacking, can directly constrain the
evolution of the $Y - M$ relation.

{\bf Contamination from radio and infrared galaxies:}
Both radio and infrared galaxies could potentially fill in SZ decrements at
150 GHz. While the contamination from radio galaxies should be small (see
discussion in Sehgal et al 2010a), the contamination from dusty star-forming
(infrared) galaxies is less clear. In Lueker et al. (2010), the signal from
infrared galaxies was removed from maps at 150 GHz by subtracting maps
at 220 GHz after fitting for a weighting factor. If all infrared sources
have the same spectral index of $\alpha = 3.6$, then this should effectively
remove infrared contamination from the 150 GHz maps. If some infrared sources
have a shallower slope (e.g., $\alpha = 2.6$ as in Knox et al. 2004), then
residual contamination will remain. However, a more recent analysis
by Shirokoff et al. (2010) suggests that even a large correlation between
infrared galaxies and groups/clusters would
not increase the 95\% CL upper limit on the thermal SZ power spectrum to the
level that it is consistent with predictions prior to Lueker et al (2010). \\

This work shows that the local groups from the S09 sample follow the extrapolation of the
pressure scaling relations from A10. More data are required to reduce the
statistical errors and more importantly explore the systematic uncertainties discussed above.
Regarding the low SZ power measured by recent experiments, we suggest some
astrophysical possibilities that may alleviate the apparent tension between
models and measurements.
Understanding the SZ power spectrum will provide important insights into both
baryon physics and cosmology.

\acknowledgments

We thank S. Allen, N. Battaglia, P. Bode, G. Holder, J. Hughes, D. Nagai, J. Ostriker, L. Shaw, J. Sievers and
H. Trac for helpful discussions.
We thank N. Battaglia, L. Shaw and H. Trac for providing their best-fits of the $Y - M$
relations.
M.S. and C.S. were supported in part by \chandra\ grants GO9-0135X and GO9-0148X, and \xmm\ grant NNX09AQ01G.
M.S., M.V. and M.D. were supported in part by the NASA LTSA grant NNG-05GD82G.
N.S. is supported by the U.S. Department of Energy contract to SLAC no. DE-AC3-76SF00515.

\vspace{-1cm}
\begin{figure}[t]
\begin{center}
\includegraphics[height=0.45\linewidth,angle=270]{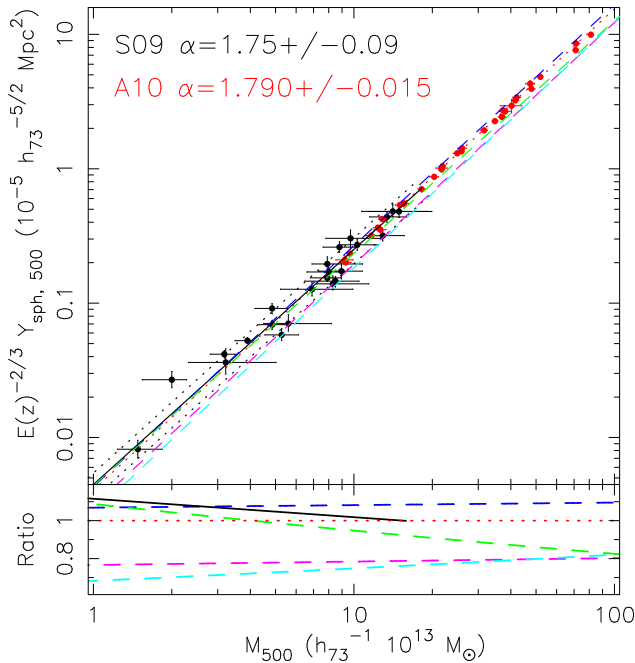}
  \caption{{\em Upper panel:} the $Y_{\rm sph, 500} - M_{500}$ relation for the S09 groups
(black points) and the A10 clusters (red points).
The best-fit relation for the S09 groups (the black solid line) is derived from the BCES
orthogonal regression method with bootstrap resampling (Akritas \& Bershady 1996), and is
given by $E(z)^{-2/3} Y_{\rm sph, 500} = 10^{\beta}$
($M_{500}$ / 3$\times10^{14}$ h$_{73}^{-1}$ M$_{\odot}$)$^{\alpha}$ h$_{73}^{-5/2}$ Mpc$^{2}$,
where $\alpha=1.75\pm0.09$ and $\beta=-4.77\pm0.09$.
The black dotted lines show the 1$\sigma$ error (22\%).
The red dotted line shows the A10 best-fit relation.
The hydrostatic equilibrium (HSE) mass values for the REXCESS clusters are not
published so the $M_{500} - Y_{\rm X, 500}$ relation (Equ.~2 in A10) was used
to derive $M_{500}$, which is the reason for the small errors and scatter of red points
as $Y_{\rm X, 500}$ and $Y_{\rm sph, 500}$ are well correlated.
Good agreement between the S09 and A10 results can be seen.
We also plot the best-fit relations from recent SZ templates;
the blue dashed line is from Sehgal et al. (2010a) ($\alpha=1.80$, $\beta=4.72$),
the green dashed line is for the AGN feedback simulations at $z=0$ by Battaglia et al. (2010)
($\alpha=1.73$, $\beta=4.81$), the magenta dashed line is from Shaw et al. (2010) at $z=0.05$
($\alpha=1.81$, $\beta=4.76$ for the HSE mass, $\alpha=1.80$, $\beta=4.85$ for the true mass)
and the cyan dashed line is for the nonthermal20 model by Trac et al. (2010)
($\alpha=1.83$, $\beta=4.86$).
We emphasize that $M_{500}$ from the X-ray data is the HSE mass which may be
smaller than the true $M_{500}$.
The models by Shaw et al. (2010)
and Trac et al. (2010) assume about 20\% non-thermal pressure support. If plotted with the
HSE mass, these two lines will shift $\sim$ 12\% to the left for $M_{500}$
(or $\sim$ 23\% higher for $Y_{\rm sph, 500}$).
{\em Lower panel:} the ratios between the SZ templates and the A10 best-fit
(the same color code as in the upper panel), while the black solid line shows
the ratio between the S09 and the A10 best-fits.
}
\end{center}
\end{figure}

\begin{figure}
\vspace{-1cm}
\centerline{\includegraphics[height=0.9\linewidth,angle=270]{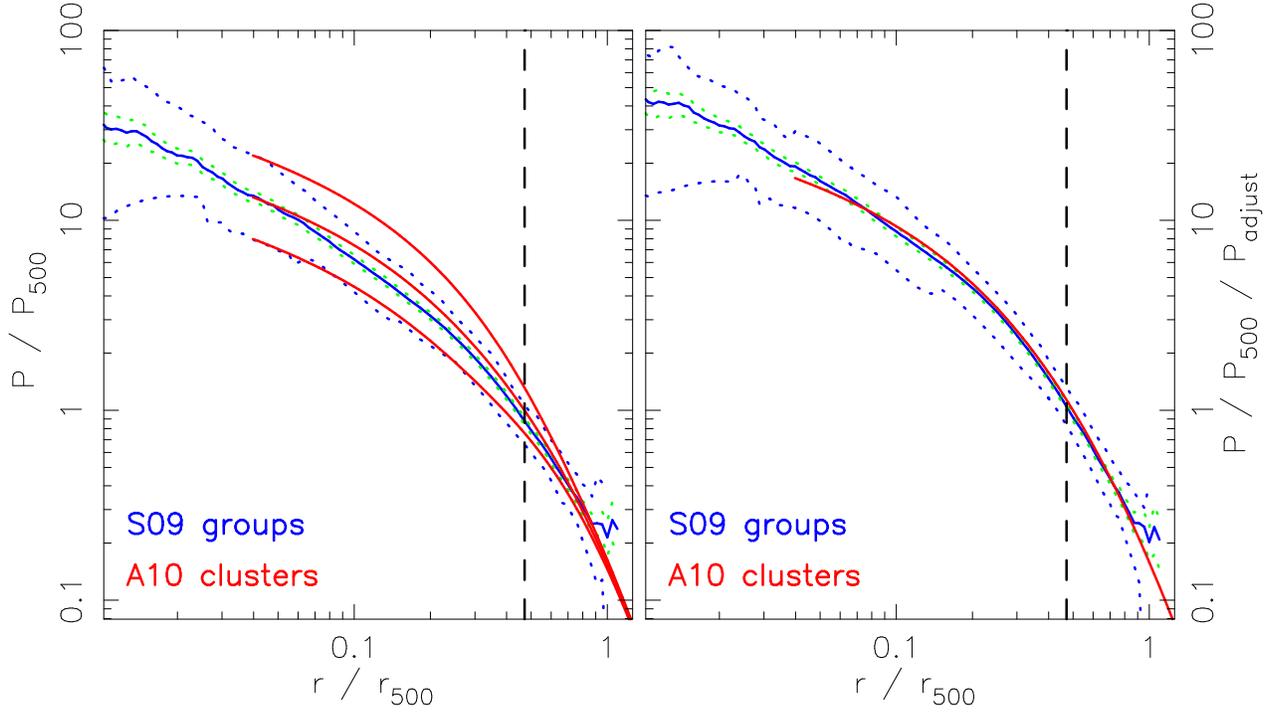}}
  \caption{Normalized pressure profiles of 43 groups from S09. $P_{500}$
is the normalization factor to remove the mass dependence if the ICM scaling
relations are self-similar (Equ.~5 of A10), while $P_{\rm adjust}$ is the
factor defined in A10 (Equ.~7 and 8) to further remove the mass dependence
as the observed ICM relations deviate from self-similarity.
The blue solid and dotted lines show the median and 1$\sigma$
scatter of the group pressure profiles. The green dotted
lines show the 1$\sigma$ uncertainties of the median.
The three red lines in the left panel are the
universal pressure profile from A10 for $M_{500} = 10^{15}$ h$_{73}^{-1}$ M$_{\odot}$,
$10^{14}$ h$_{73}^{-1}$ M$_{\odot}$ and $10^{13}$ h$_{73}^{-1}$ M$_{\odot}$,
from the top to the bottom respectively. The median $M_{500}$ of the S09
groups is $\sim 7\times10^{13}$ h$_{73}^{-1}$ M$_{\odot}$.
The red line in the right panel is the universal pressure profile from A10
after removing the mass dependence. The median pressure profile for the
S09 groups agrees well with the A10 profile.
The dashed line shows the position of $r_{2500}$.
}
\end{figure}


\begin{references}

 \reference{} Akritas, M. G., \& Bershady, M. A. 1996, ApJ, 470, 706
 \reference{} Arnaud, M., Pointecouteau, E., \& Pratt, G. W. 2007, A\&A, 474, L37
 \reference{} Arnaud, M. et al. 2010, A\&A, 517, 92 (A10)
 \reference{} Battaglia, N. et al. 2010, ApJ, 725, 91
 \reference{} Bode, P., Ostriker, J. P., \& Vikhlinin, A. 2009, ApJ, 700, 989
 \reference{} Burns, J. O. et al. 2010, ApJ, 721, 1105
 \reference{} Dai, X. et al. 2010, ApJ, 719, 119
 \reference{} Das, S. et al.~2010, arXiv:1009.0847
 \reference{} Dunkley, J. et al. 2010, arXiv:1009.0866
 \reference{} Finoguenov, A. et al. 2007, MNRAS, 374, 737
 \reference{} Jeltema, T. E. et al. 2009, MNRAS, 399, 715
 \reference{} Knox, L., Holder, G. P., \& Church, S. E. 2004, ApJ, 612, 96
 \reference{} Komatsu, E., \& Seljak, U. 2002, MNRAS, 336, 1256
 \reference{} Komatsu, E. et al. 2010, ApJS, in press, arXiv:1001.4538
 \reference{} Lau, E. T., Kravtsov, A. V., \& Nagai, D. 2009, ApJ, 705, 1129
 \reference{} Leauthaud, A. et al. 2010, ApJ, 709, 97
 \reference{} Lueker, M. et al. 2010, ApJ, 719, 1045
 \reference{} Mahdavi, A. et al. 2005, ApJ, 622, 187
 \reference{} McCarthy, I. G. et al. 2010, MNRAS, in press, arXiv:1008.4799
 \reference{} Nagai, D. et al. 2007, ApJ, 668, 1
 \reference{} Nevalainen, J., David, L., \& Guainazzi, M. 2010, A\&A, 523, 22
 \reference{} Pacaud, F. et al. 2007, MNRAS, 382, 1289
 \reference{} Plagge, T. et al. 2010, ApJ, 716, 1118
 \reference{} Ponman, T. J. et al. 2003, MNRAS, 343, 331
 \reference{} Popesso, P. et al. 2007, A\&A, 461, 397
 \reference{} Pratt, G. W. et al. 2010, A\&A, 511, 85
 \reference{} Rasmussen, J. et al. 2006, MNRAS, 373, 653
 \reference{} Rines, K., \& Diaferio, A. 2010, AJ, 139, 580
 \reference{} Rykoff, E. S. et al. 2008, ApJ, 675, 1106
 \reference{} Sehgal, N. et al. 2010a, ApJ, 709, 920
 \reference{} Sehgal, N. et al. 2010b, ApJ submitted, arXiv:1010.1025
 \reference{} Shaw, L. et al. 2010, ApJ, arXiv:1006.1945	
 \reference{} Shirokoff, E. et al. 2010, ApJ, submitted, arXiv:1012.4788
 \reference{} Sun, M. et al. 2009, ApJ, 693, 1142 (S09)
 \reference{} Trac, H., Bode, P., \& Ostriker, J. P. 2010, arXiv1006.2828
 \reference{} Vikhlinin, A. et al. 2006, ApJ, 640, 691
 \reference{} Vikhlinin, A. et al. 2009, ApJ, 692, 1033
 \reference{} Voit, G. M. 2005, Rev. Mod. Phys., 77, 207

\end{references}
\end{document}